\documentclass[aps,pra,twocolumn,reprint]{revtex4-2}
\usepackage[utf8]{inputenc}
\usepackage[T1]{fontenc}
\usepackage{amsmath,amssymb}
\usepackage{graphicx}
\usepackage[colorlinks,citecolor=blue,urlcolor=blue,linkcolor=blue,breaklinks]{hyperref}
\usepackage{tikz}

\begin{document}

\title{Beyond the Lorenz Gauge: Probing a Stueckelberg Scalar in the Electric Aharonov-Bohm Effect}

\author{Renato Vieira dos Santos}
\email{renato.santos@ufla.br}
\affiliation{Instituto de Ci\^{e}ncia, Tecnologia e Inova\c{c}\~{a}o -- ICTIN, Universidade Federal de Lavras -- UFLA, Campus Para\'{i}so, MG 37950-000, Brazil}

\begin{abstract}
The electric Aharonov-Bohm effect---a time-dependent scalar potential imparting a measurable phase shift on electrons in a region free of electromagnetic fields---has never been experimentally tested in its original formulation with shielded, time-dependent potentials. This unexplored regime offers a rare opportunity: the Lorenz condition $\partial_\mu A^\mu = 0$, a choice that eliminates a scalar degree of freedom from the electromagnetic potential, may not be the last word. We consider a gauge-invariant extension of quantum electrodynamics in which a Stueckelberg scalar field survives as a physical degree of freedom and couples to electrons. In the shielded electrostatic AB configuration, this coupling yields a phase shift with a distinctive $1-\cos(\omega T)$ signature---orthogonal to the standard $\sin(\omega T)$ and separable by a frequency sweep even if both contributions coexist. We propose a measurement protocol based on single-electron interferometry with picosecond time resolution, within reach of current technology. The experiment asks a question that has lingered since 1959: is the Lorenz gauge a matter of convenience, or a matter of principle?
\end{abstract}

\maketitle

\section{Introduction}
\label{sec:intro}

In 1959, Aharonov and Bohm \cite{aharonov1959} predicted that an electron passing through a region where the electric and magnetic fields are strictly zero can experience a measurable phase shift, provided the electromagnetic potentials are non-zero in that region. The magnetic version of the Aharonov-Bohm (AB) effect has been confirmed with extraordinary precision~\cite{tonomura1986, webb1985, batelaan2009}. Its topological protection by the non-simple connectivity of the electron's configuration space~\cite{wu1975} places it among the most robust predictions of quantum mechanics.

The electric version is more subtle. A time-dependent scalar potential is applied to conducting tubes through which the electron passes, while the electric field inside is maintained at zero by electrostatic shielding~\cite{aharonov1959}. Standard quantum mechanics, through minimal coupling, predicts a phase shift proportional to the time integral of the potential: $\Delta\phi_{\mathrm{QM}} = -(e/\hbar) \int \Delta\Phi(t) \, dt$. However, the \emph{functional dependence} of the phase on the temporal profile of the applied potential has never been systematically mapped~\cite{batelaan2009}. For a static potential, the phase grows linearly with the transit time. For a time-varying potential, the exact dependence on frequency and waveform remains experimentally unexplored.

This gap is significant. The only experimental realization of the electric AB configuration to date is the steady-state Matteucci-Pozzi (MP) effect~\cite{matteucci1985}. However, Hilbert \emph{et al.}~\cite{hilbert2011} demonstrated that the MP phase shift is accompanied by a classical time delay and is therefore a force-based effect, not a genuine AB phenomenon. The original electric AB configuration---with time-dependent potentials applied to shielded cylinders---has never been tested.

This unexplored regime invites the question: could the electron's phase receive contributions beyond minimal coupling when the electric field is strictly zero? The electromagnetic potential $A_\mu = (\Phi/c, \mathbf{A})$ has four components. Two correspond to the transverse photon polarizations; the other two are eliminated in standard QED by gauge fixing and the Lorenz condition $\partial_\mu A^\mu = 0$. The Lorenz condition is, however, a \emph{choice}, not a dynamical law. In 1938, Stueckelberg~\cite{stueckelberg1938} introduced a compensating scalar field that preserves gauge invariance while allowing the longitudinal scalar mode to become physical. The scalar quantity $B = \partial_\mu A^\mu$, traditionally eliminated by gauge fixing, can survive as a gauge-invariant operator when paired with the Stueckelberg field. Although the Stueckelberg mechanism was originally devised for massive electrodynamics, it can be adapted to massless QED as well: the Stueckelberg field provides a gauge-invariant way to isolate the longitudinal mode without imposing the Lorenz condition as a fundamental requirement.

In this framework, the most general interaction Lagrangian between a Dirac field $\psi$ and the electromagnetic sector that respects the full $U(1)$ gauge symmetry can include gauge-invariant scalar operators built from $B$. To leading order in a derivative expansion, the relevant terms are
\begin{equation}
\mathcal{L}_{\text{int}} = -e \bar{\psi}\gamma^\mu\psi \tilde{A}_\mu - \frac{g}{\Lambda} \bar{\psi}\psi \, B + \frac{g'}{\Lambda} \bar{\psi}\gamma^5\psi \, B,
\label{eq:L_int}
\end{equation}
where $\tilde{A}_\mu = A_\mu + \partial_\mu \phi$ is the gauge-invariant Stueckelberg combination, $\phi$ is the Stueckelberg scalar field, $B = \partial_\mu \tilde{A}^\mu$ is the gauge-invariant scalar operator, $\Lambda$ is a heavy scale (at least at the electroweak scale), and $g$, $g'$ are dimensionless Wilson coefficients. The first term reproduces the standard minimal coupling because $\tilde{A}_\mu$ differs from $A_\mu$ by a pure derivative that integrates to zero in the phase. The second and third are Pauli-like scalar and pseudoscalar couplings, respectively. They are Lorentz-invariant, gauge-invariant, and absent from standard QED only because the Lorenz condition is imposed (which corresponds to the limit where the Stueckelberg field decouples). Their effects are suppressed by the heavy scale $\Lambda$ and vanish in the static limit where all existing precision tests operate---precisely the properties of the coupling proposed here.

In the non-relativistic limit, the pseudoscalar term is spin-suppressed, and the scalar term contributes an effective potential $V_{\text{eff}} = (g/\Lambda) B$ that adds to the electron's phase. In the shielded electrostatic AB configuration with ideal conductors, $\mathbf{A} = 0$ and $\mathbf{E} = 0$ inside the drift tubes. The Stueckelberg field $\phi$ can be chosen to vanish on the inner surfaces of the cylinders by a residual gauge transformation, giving
\begin{equation}
B = \partial_\mu \tilde{A}^\mu = \nabla\cdot\mathbf{A} + \frac{1}{c^2}\frac{\partial\Phi}{\partial t} = \frac{\dot{\Phi}}{c^2}.
\label{eq:B_shielded}
\end{equation}
The phase shift from the scalar coupling becomes
\begin{equation}
\Delta\phi_{B} = -\frac{g}{\hbar\Lambda} \int_0^T \Delta\dot{\Phi}(t) \, dt = -\frac{g}{\hbar\Lambda} \bigl[ \Delta\Phi(T) - \Delta\Phi(0) \bigr].
\label{eq:phase_B}
\end{equation}
This is a boundary term: only the initial and final values of the potential difference matter. For a sinusoidal modulation $\Delta\Phi(t) = \Delta\Phi_0 \cos(\omega t)$, the standard and scalar-coupling predictions are
\begin{align}
\Delta\phi_{\mathrm{QM}} &= -\frac{e\Delta\Phi_0}{\hbar\omega} \sin(\omega T), \label{eq:sin_intro} \\
\Delta\phi_{B} &= -\frac{g\Delta\Phi_0}{\hbar\Lambda} \bigl[1 - \cos(\omega T)\bigr], \label{eq:cos_intro}
\end{align}
These two functions are linearly independent: their zero-crossings occur at $\omega T = n\pi$ versus $2n\pi$, their maxima at different values, and their low-frequency limits are linear and quadratic in $\omega T$, respectively. If both couplings coexist, the total phase is $A\sin(\omega T) + B[1-\cos(\omega T)]$, and a frequency sweep cleanly separates them.

The primary goal of this paper is to demonstrate that this separation is experimentally feasible with current technology. We proceed as follows. Section~\ref{sec:theory} reviews the gauge-invariant Stueckelberg formalism and the phenomenological coupling structure. Section~\ref{sec:prediction} formalizes the phase prediction and its experimental separability. Section~\ref{sec:experiment} outlines a measurement protocol based on single-electron interferometry with picosecond time resolution. Section~\ref{sec:discussion} addresses the relationship with the Josephson effect and other precision tests. Section~\ref{sec:conclusion} summarizes our findings.

\section{Gauge-Invariant Stueckelberg Scalar and Its Couplings}
\label{sec:theory}

\subsection{Gauge invariance and the Stueckelberg scalar}

In standard electrodynamics, the field strength tensor $F_{\mu\nu} = \partial_\mu A_\nu - \partial_\nu A_\mu$ is invariant under local gauge transformations $A_\mu \to A_\mu + \partial_\mu \chi$. The four-vector potential $A_\mu = (\Phi/c, \mathbf{A})$ has four components. Two of them are eliminated by gauge fixing: one via a gauge constraint (e.g., $A_0 = 0$ in Coulomb gauge), and another via the Lorenz condition $\partial_\mu A^\mu = 0$, which removes the longitudinal scalar mode~\cite{weinberg1995}. This leaves the two transverse photon polarizations as the only physical states.

The Lorenz condition is, however, a choice, not a dynamical requirement. To promote the longitudinal scalar to a physical degree of freedom while preserving exact $U(1)$ gauge invariance, we introduce a Stueckelberg scalar field $\phi$ with the transformation rule
\begin{equation}
\phi \to \phi + \chi \quad \text{when} \quad A_\mu \to A_\mu + \partial_\mu \chi.
\label{eq:phi_transf}
\end{equation}
The combination
\begin{equation}
\tilde{A}_\mu \equiv A_\mu + \partial_\mu \phi
\label{eq:tildeA}
\end{equation}
is then gauge-invariant. Its field strength is identical to $F_{\mu\nu}$: $\partial_\mu \tilde{A}_\nu - \partial_\nu \tilde{A}_\mu = F_{\mu\nu}$. The four-divergence
\begin{equation}
B \equiv \partial_\mu \tilde{A}^\mu = \partial_\mu (A^\mu + \partial^\mu \phi)
\label{eq:B_def}
\end{equation}
is also gauge-invariant. We refer to $B$ as the Stueckelberg scalar. When $\phi$ possesses its own kinetic term, $B$ becomes a propagating physical field. In the limit where the kinetic term for $\phi$ is removed (i.e., $\phi$ becomes an auxiliary field), its equation of motion enforces $B = 0$, and we recover standard QED with the Lorenz condition as an on-shell constraint. The Stueckelberg formalism thus provides a continuous interpolation between standard QED and a theory with a physical longitudinal scalar mode.

The historical roots of this idea trace back to Stueckelberg~\cite{stueckelberg1938}, who introduced the compensating field to preserve gauge invariance in massive electrodynamics. Fock and Podolsky~\cite{fock1932}, Dirac, Fock, and Podolsky~\cite{dirac1932}, and Ohmura~\cite{ohmura1956} explored formulations where the Lorenz scalar plays a dynamical role. For a modern treatment of the Stueckelberg mechanism, see Ref.~\cite{ruegg2004}. More recently, Modanese~\cite{modanese2017, modanese2023b} has investigated generalized Maxwell equations where the Lorenz condition is promoted to a dynamical equation for gauge waves, an approach that naturally accommodates a physical $B$ field.

\subsection{Phenomenological effective couplings}

With the gauge-invariant building blocks $\tilde{A}_\mu$, $B$, and $F_{\mu\nu}$, we can construct the most general interaction Lagrangian between a Dirac field $\psi$ and the gauge sector that respects Lorentz invariance and the full $U(1)$ gauge symmetry. To leading order in a derivative expansion, the relevant terms are given by Eq.~\eqref{eq:L_int}. The first term, $-e\bar{\psi}\gamma^\mu\psi \tilde{A}_\mu$, reproduces the standard minimal coupling: the difference between $\tilde{A}_\mu$ and $A_\mu$ is a pure derivative that integrates to zero in the action. The second and third terms are scalar and pseudoscalar couplings to $B$, respectively. They are gauge-invariant and Lorentz-invariant operators of dimension 4 (since $B$ has dimension 2 and $1/\Lambda$ compensates).

We focus on the scalar term because the pseudoscalar term is spin-suppressed in the non-relativistic limit relevant for electron interferometry.\footnote{In the non-relativistic limit, the pseudoscalar bilinear $\bar{\psi}\gamma^5\psi$ couples to the electron spin operator $\boldsymbol{\sigma}\cdot\mathbf{p}/m$ rather than to the charge density $\rho = \psi^\dagger\psi$. For unpolarized electron beams, as used in typical interferometry setups, the spin-dependent contribution averages to zero over the ensemble. Even for a polarized beam, the suppression is of order $v/c \sim 10^{-2}$ for 100~eV electrons relative to the scalar term. A dedicated experiment with spin-polarized electrons could in principle separate the two contributions, but this is beyond the scope of the present proposal.} The total Lagrangian is
\begin{equation}
\mathcal{L} = \bar{\psi}(i\gamma^\mu\partial_\mu - m)\psi - \frac{1}{4}F_{\mu\nu}F^{\mu\nu} - e\bar{\psi}\gamma^\mu\psi \tilde{A}_\mu - \frac{g}{\Lambda}\bar{\psi}\psi B + \mathcal{L}_{\phi},
\label{eq:L_total}
\end{equation}
where $\mathcal{L}_{\phi}$ contains the kinetic and potential terms for the Stueckelberg field $\phi$. In the minimal scenario where $\phi$ is an auxiliary field with no independent dynamics, $\mathcal{L}_{\phi}$ consists of a topological term or is simply absent; the Euler-Lagrange equation for $\phi$ then enforces $\partial_\mu(\bar{\psi}\psi \partial^\mu \phi) = 0$, which in the shielded AB configuration reduces to a boundary condition. The scale $\Lambda$ is unknown but is constrained to be at least at the electroweak scale by the absence of observable scalar electrodynamics in collider experiments; $g$ is a Wilson coefficient of order unity if the coupling originates from physics at scale $\Lambda$.\footnote{Laboratory constraints on new scalar-mediated forces of range $\lambda$ set limits on the coupling constant as a function of mass $m = \hbar/(c\lambda)$. For a massless scalar (infinite range), torsion-balance experiments constrain the coupling to matter to be below $10^{-21}$ of gravitational strength~\cite{adelberger2009}, which for a coupling of the form $g/\Lambda$ translates to $g/\Lambda \lesssim 10^{-19}$~GeV$^{-1}$, consistent with $\Lambda \gtrsim 1$~TeV for $g \sim \mathcal{O}(1)$. The experiment proposed here probes a different regime---time-dependent, field-free---and is complementary to static force searches.}

\subsection{Non-relativistic limit and effective phase}

In the non-relativistic limit, the Dirac bilinear $\bar{\psi}\psi$ reduces to the probability density $\psi^\dagger\psi$ up to corrections of order $v^2/c^2$. The interaction Hamiltonian density becomes
\begin{equation}
\mathcal{H}_{\text{int}} = e\Phi \, \psi^\dagger\psi + \frac{g}{\Lambda}B \, \psi^\dagger\psi.
\label{eq:H_int}
\end{equation}
For a single-electron wave packet traversing a region where $\Phi$ and $B$ vary slowly compared to the de Broglie wavelength, the eikonal approximation gives the phase accumulated along a classical trajectory:
\begin{equation}
\Delta\phi = -\frac{1}{\hbar}\int_0^T \left[ e\Delta\Phi(t) + \frac{g}{\Lambda}\Delta B(t) \right] dt.
\label{eq:total_phase}
\end{equation}
The first term is the standard AB phase. The second is the contribution from the Stueckelberg scalar.

In the shielded electrostatic AB configuration, the conducting drift tubes enforce $\mathbf{E} = 0$ and $\mathbf{A} = 0$ in their interior. Maxwell's equations in vacuum, $\nabla\cdot\mathbf{E} = 0$, are satisfied trivially. The Stueckelberg field $\phi$ can be chosen to vanish on the inner surfaces of the cylinders by a residual gauge transformation (since $\phi$ shifts by the same function as $A_\mu$ under gauge transformations). With this choice, $\tilde{A}_\mu = A_\mu$ inside the tubes, and the scalar field $B$ reduces to the kinematic identity $B = \dot{\Phi}/c^2$. Substituting into Eq.~\eqref{eq:total_phase} gives
\begin{equation}
\Delta\phi = -\frac{e}{\hbar}\int_0^T \Delta\Phi(t)\,dt - \frac{g}{\hbar\Lambda c^2} \bigl[\Delta\Phi(T) - \Delta\Phi(0)\bigr].
\label{eq:phase_combined}
\end{equation}
The scalar contribution collapses to a boundary term: it depends only on the net change in the potential difference, not on its time integral. For a static potential, $\Delta\Phi(T) = \Delta\Phi(0)$ and the scalar contribution vanishes identically.\footnote{The conditions $\mathbf{E}=0$ and $\mathbf{A}=0$ inside an ideal conducting cylinder with a time-dependent but spatially uniform surface potential can be established as follows. In the Coulomb gauge ($\nabla\cdot\mathbf{A}=0$), the electric field is $\mathbf{E} = -\nabla\Phi - \partial_t\mathbf{A}$. For a cylindrical conductor held at potential $\Phi(t)$ on its inner surface, the Laplace equation $\nabla^2\Phi = 0$ with Dirichlet boundary condition $\Phi|_{\text{surface}} = \Phi(t)$ has the unique solution $\Phi(\mathbf{r},t) = \Phi(t)$ throughout the interior, giving $\nabla\Phi = 0$. Maxwell's equations in vacuum then give $\Box\mathbf{A} = 0$. With the boundary condition $\mathbf{A} = 0$ on the inner surface (the tangential component vanishes for a perfect conductor, and the normal component can be gauged to zero), the unique solution is $\mathbf{A} = 0$ everywhere inside. Hence $\mathbf{E} = 0$, $\mathbf{A} = 0$, and $B = \nabla\cdot\mathbf{A} + \dot{\Phi}/c^2 = \dot{\Phi}/c^2$ exactly. Corrections from finite conductivity are discussed in Sec.~\ref{sec:experiment}. The Stueckelberg field $\phi$ enters through $\tilde{A}_\mu = A_\mu + \partial_\mu \phi$; choosing $\phi = 0$ on the inner surface by a residual gauge transformation ensures that $\tilde{A}_\mu$ satisfies the same boundary conditions as $A_\mu$, preserving the field-free condition inside the tubes.}

\section{Prediction for the Electric Aharonov-Bohm Effect}
\label{sec:prediction}

\subsection{Standard quantum mechanical phase}
\label{sec:standard}

We consider the electric AB configuration: a coherent electron beam is split into two paths that pass through separate conducting tubes to which potentials $\Phi_1(t)$ and $\Phi_2(t)$ are applied. Potentials are spatially uniform within each tube. The electron, of charge $-e$, travels through each tube during a time interval $T$. In standard quantum mechanics, the phase difference between the two paths is
\begin{equation}
\Delta \phi_{\mathrm{QM}} = -\frac{e}{\hbar} \int_0^T \bigl[ \Phi_1(t) - \Phi_2(t) \bigr] \, dt.
\label{eq:QM_phase}
\end{equation}
For a static potential, $\Delta\phi_{\mathrm{QM}} = -(e/\hbar) \Delta\Phi \, T$, linear in the transit time.

\subsection{Scalar coupling contribution}
\label{sec:B_phase}

From Eq.~\eqref{eq:phase_combined}, the scalar coupling contribution to the phase difference is
\begin{equation}
\Delta \phi_{B} = -\frac{g}{\hbar\Lambda c^2} \Bigl[ \Delta\Phi(T) - \Delta\Phi(0) \Bigr].
\label{eq:B_phase_final}
\end{equation}
The entire intermediate history of $\Delta\Phi(t)$ is irrelevant. This is a direct consequence of the fundamental theorem of calculus applied to the time derivative of the potential.

Several features deserve emphasis. For a static potential, $\Delta\Phi(T) = \Delta\Phi(0)$ and $\Delta\phi_B = 0$. A constant potential difference produces no phase shift from this coupling, regardless of magnitude or transit time. The DC phase shifts observed in electron interferometry---for example, from electrostatic biprism potentials---are force-based effects arising from non-zero electric fields. They do not test the field-free AB configuration.

The phase is bounded for any physically realizable modulation, since $\Delta\Phi$ cannot diverge. In contrast, the standard phase for a static potential grows without bound as $T$ increases.

\subsection{Coexistence and experimental separability}
\label{sec:orthogonality}

If both the standard minimal coupling and the scalar coupling are operative, the total phase difference is
\begin{equation}
\Delta\phi = -\frac{e}{\hbar}\int_0^T \Delta\Phi(t)\,dt - \frac{\kappa}{\hbar} \bigl[\Delta\Phi(T) - \Delta\Phi(0)\bigr],
\label{eq:combined}
\end{equation}
where $\kappa = g/(\Lambda c^2)$. The two terms are functionally orthogonal for any non-trivial time dependence. The integral $\int \Delta\Phi \, dt$ is sensitive to the entire history; the boundary term $\Delta\Phi(T) - \Delta\Phi(0)$ depends only on the endpoints. A measurement that varies the modulation parameters can separate them unambiguously.

\subsection{Sinusoidal modulation}
\label{sec:sinusoidal}

For a sinusoidal modulation of the potential difference,
\begin{equation}
\Delta\Phi(t) = \Delta\Phi_0 \cos(\omega t),
\label{eq:sin_mod}
\end{equation}
the electron enters at $t=0$ and exits at $t=T$. The standard phase, from Eq.~\eqref{eq:QM_phase}, is
\begin{equation}
\Delta \phi_{\mathrm{QM}} = -\frac{e \Delta\Phi_0}{\hbar \omega} \sin(\omega T).
\label{eq:QM_sin}
\end{equation}
In the low-frequency regime $\omega T \ll 1$, this reduces to $\Delta\phi_{\mathrm{QM}} \approx -(e\Delta\Phi_0/\hbar) T$, linear in the transit time.

The scalar coupling phase, from Eq.~\eqref{eq:B_phase_final}, is
\begin{equation}
\Delta \phi_{B} = -\frac{\kappa \Delta\Phi_0}{\hbar} \Bigl[ 1 - \cos(\omega T) \Bigr].
\label{eq:B_sin}
\end{equation}
In the low-frequency limit,
\begin{equation}
\Delta \phi_{B} \approx -\frac{\kappa \Delta\Phi_0}{2\hbar} (\omega T)^2,
\label{eq:B_approx}
\end{equation}
quadratic in both frequency and transit time. We note that the coefficient $\kappa/\hbar = g/(\hbar\Lambda c^2)$ is suppressed by the heavy scale $\Lambda$, which is at least at the electroweak scale ($\Lambda \gtrsim 1$~TeV). Numerically, $e/\hbar \approx 1.52 \times 10^{15}$~(V$\cdot$s)$^{-1}$, while $\kappa/\hbar \approx 2.4 \times 10^{-23}\,g$~(V$\cdot$s)$^{-1}$ for $\Lambda = 1$~TeV. For $g$ of order unity, the scalar coupling phase is many orders of magnitude smaller than the standard phase for the same applied potential. The experiment therefore tests the \emph{functional form} $1-\cos(\omega T)$, not a specific numerical prediction for the phase magnitude. The coefficient of the $1-\cos(\omega T)$ term is treated as a free parameter to be extracted from the data, as discussed in Sec.~\ref{sec:protocol}. A non-zero value would indicate the presence of the scalar coupling, regardless of its absolute magnitude.

To estimate the sensitivity of the proposed experiment, consider a frequency sweep with $N_\omega$ points, each measured with a phase uncertainty $\sigma_\phi$. The statistical uncertainty on the coefficient $B$ scales as $\sigma_B \sim \sigma_\phi / \sqrt{N_\omega}$. For $\sigma_\phi \approx 10^{-2}$~rad and $N_\omega \approx 50$, this gives $\sigma_B \approx 1.4 \times 10^{-3}$~rad, corresponding to a $1\sigma$ upper limit on the coupling of $\kappa/\hbar \lesssim 1.4 \times 10^{-3}$~(V$\cdot$s)$^{-1}$. While this is less sensitive than static fifth-force searches for massless scalars, it probes a qualitatively different regime---time-dependent, field-free---where existing bounds do not apply. For comparison, the standard minimal coupling is $e/\hbar \approx 1.52 \times 10^{15}$~(V$\cdot$s)$^{-1}$; the proposed experiment would thus be sensitive to a scalar coupling approximately $10^{18}$ times weaker than the electromagnetic interaction, illustrating the remarkable precision achievable with modern electron interferometry~\cite{hasselbach2010}.

\begin{figure}[!htb]
\centering
\begin{tikzpicture}[scale=0.9]
  \draw[->] (-0.3,0) -- (7.5,0) node[right] {$\omega T$};
  \draw[->] (0,-2.5) -- (0,2.5) node[above] {$\Delta\phi$ (arb. units)};
  \draw[domain=0:7,smooth,variable=\x,blue,thick] plot ({\x},{sin(\x r)});
  \draw[domain=0:7,smooth,variable=\x,red,thick] plot ({\x},{1-cos(\x r)});
  \node[blue,above] at (4.7,-1.55) {$\sin(\omega T)$};
  \node[red,below] at (5.0,2) {$1-\cos(\omega T)$};
  \draw[dotted] (3.14,-2.5) -- (3.14,2.5);
  \draw[dotted] (6.28,-2.5) -- (6.28,2.5);
  \node at (3.14,-2.7) {$\pi$};
  \node at (6.28,-2.7) {$2\pi$};
\end{tikzpicture}
\caption{Standard ($\sin\omega T$, blue) and scalar coupling ($1-\cos\omega T$, red) phase predictions for sinusoidal modulation. Zero-crossings occur at $\omega T = n\pi$ (standard) and $\omega T = 2n\pi$ (scalar coupling). The two functions are linearly independent, enabling separation in a combined fit $A\sin(\omega T) + B[1-\cos(\omega T)]$. Both curves are shown in arbitrary units; their relative normalization is determined by the fit coefficients $A$ and $B$ in Eq.~\eqref{eq:fit}.}
\label{fig:comparison}
\end{figure}
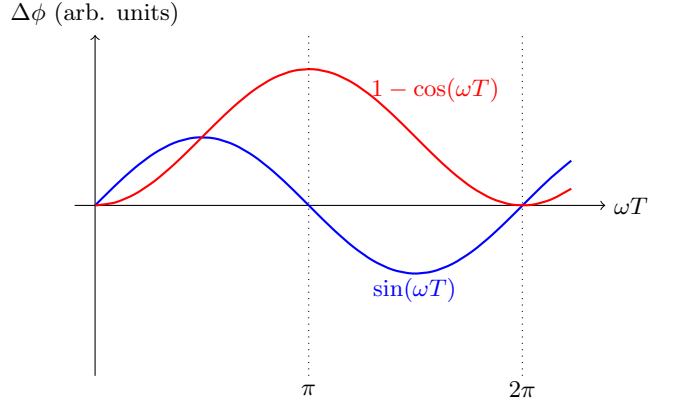

The contrast between the two predictions is summarized in Table~\ref{tab:comparison} and illustrated in Fig.~\ref{fig:comparison}. The functional forms $\sin(\omega T)$ and $1 - \cos(\omega T)$ are qualitatively different in their dependence on $\omega T$. Their zero-crossings occur at distinct, non-overlapping values ($\omega T = n\pi$ versus $2n\pi$), and their maxima at distinct values ($(2n+1)\pi/2$ versus $(2n+1)\pi$). This linear independence ensures that a combined fit $A\sin(\omega T) + B[1-\cos(\omega T)]$ can resolve the two components.

\begin{table*}[!htb]
\centering
\caption{Comparison of phase shift predictions for a sinusoidal potential modulation $\Delta\Phi(t) = \Delta\Phi_0 \cos(\omega t)$. The electron transits the modulated region in time $T$.}
\begin{tabular}{|l|c|c|}
\hline
\textbf{Feature} & \textbf{Standard QM} & \textbf{Scalar coupling (this work)} \\
\hline
Static potential ($\omega = 0$) & $\Delta\phi \propto T$ & $\Delta\phi = 0$ \\
\hline
Low-$\omega T$ scaling & Linear: $\propto T$ & Quadratic: $\propto (\omega T)^2$ \\
\hline
Functional form & $-\frac{e\Delta\Phi_0}{\hbar\omega} \sin(\omega T)$ & $-\frac{\kappa\Delta\Phi_0}{\hbar}[1 - \cos(\omega T)]$ \\
\hline
Zeroes at & $\omega T = n\pi$ & $\omega T = 2n\pi$ \\
\hline
Maxima at & $\omega T = (2n+1)\pi/2$ & $\omega T = (2n+1)\pi$ \\
\hline
Maximum magnitude & $\frac{e\Delta\Phi_0}{\hbar\omega}$ & $\frac{2\kappa\Delta\Phi_0}{\hbar}$ \\
\hline
\end{tabular}
\label{tab:comparison}
\end{table*}

\section{Experimental Feasibility}
\label{sec:experiment}

\subsection{Parameter regime}
\label{sec:parameters}

Distinguishing the $\sin(\omega T)$ and $1 - \cos(\omega T)$ predictions requires access to $\omega T \sim 1$. For $\omega T \ll 1$, both are small and the difference between linear and quadratic scaling is subtle. For $\omega T \gg 1$, both oscillate and phase-averaging from finite beam energy spread may obscure the contrast.

The transit time $T$ is set by the electron velocity and the length $L$ of the modulated region. For an electron of kinetic energy $E = m_e v^2 / 2$,
\begin{equation}
v = \sqrt{\frac{2E}{m_e}} \approx 5.93 \times 10^5 \sqrt{E \, (\text{in eV})} \ \text{m/s}.
\label{eq:velocity}
\end{equation}
Table~\ref{tab:transit} lists transit times for $L = 1$~cm.

\begin{table}[!htb]
\centering
\caption{Electron transit times for a modulated region of length $L = 1$~cm.}
\begin{tabular}{|c|c|c|}
\hline
$E$~(eV) & $v$~(m/s) & $T$~(s) \\
\hline
10  & $1.88 \times 10^6$ & $5.33 \times 10^{-9}$ \\
100 & $5.93 \times 10^6$ & $1.69 \times 10^{-9}$ \\
$10^3$ & $1.88 \times 10^7$ & $5.33 \times 10^{-10}$ \\
$10^4$ & $5.93 \times 10^7$ & $1.69 \times 10^{-10}$ \\
\hline
\end{tabular}
\label{tab:transit}
\end{table}

The condition $\omega T \sim 1$ corresponds to modulation frequencies $f = \omega/2\pi \approx 30$~MHz to $1$~GHz, standard RF technology. At $f \sim 1$~GHz, the free-space wavelength $\lambda \sim 30$~cm is substantially larger than $L = 1$~cm. The round-trip time of an electromagnetic signal across the apparatus ($\sim 2L/c \approx 7 \times 10^{-11}$~s) is much shorter than the modulation period ($\sim 10^{-9}$~s), justifying the quasi-static assumption of spatial uniformity. Residual spatial variations near cylinder apertures, where fringe fields penetrate over a distance comparable to the cylinder radius, are suppressed by ensuring the electron beam passes at least several cylinder radii from the apertures. For a cylindrical tube of radius $R$, the fringe electric field near the aperture decays exponentially with distance from the opening over a characteristic scale $\ell_{\text{fringe}} \sim R$. For a beam passing at a distance $r \gtrsim 3R$ from the aperture, the residual field is suppressed by a factor $\sim e^{-3} \approx 0.05$. At 100~eV electron energy, a residual field of $10^{-3}$ times the applied potential difference over a 0.1~mm path contributes a phase of order $10^{-3}$~rad, well below the precision needed to resolve the functional signatures.

At GHz modulation frequencies, the cylindrical drift tubes may support cavity resonances if their length becomes comparable to the wavelength of the RF signal. For 1~cm tubes at 1~GHz, the length is only $\lambda/30$, so the lowest cavity modes are far below cutoff. Nevertheless, careful impedance matching at the electrode terminations is required to prevent reflections that could introduce standing-wave components in the applied potential. These are standard microwave engineering challenges that can be addressed with established techniques~\cite{hasselbach2010}.

At GHz frequencies, the finite conductivity of the drift tube walls leads to a non-zero skin depth $\delta = \sqrt{2/\mu_0\sigma\omega}$. For copper ($\sigma \approx 5.8 \times 10^7$~S/m) at 1~GHz, $\delta \approx 2$~$\mu$m. This is much smaller than typical tube wall thicknesses ($\sim 0.1$--1~mm), so the penetration of RF fields through the walls is negligible. However, the finite skin depth implies that surface currents in the conductor generate a small tangential magnetic field just outside the inner surface, which can produce a residual azimuthal vector potential $A_\phi$ inside the tube. For a tube radius $R$, the residual field is suppressed by a factor $\sim \delta/R \sim 10^{-3}$ for millimeter-scale tubes. The corresponding correction to $B = \dot{\Phi}/c^2$ is of order $\nabla\cdot\mathbf{A} \sim A_\phi/R \sim 10^{-3} \dot{\Phi}/c^2$, contributing a phase of order $10^{-3}$~rad and thus below the statistical sensitivity estimated above.

The amplitude $\Delta\Phi_0$ sets the scale. The prefactor $e/\hbar \approx 1.52 \times 10^{15}$~rad/(V$\cdot$s) means $1$~V over nanoseconds produces phases of order $10^{5}$--$10^{7}$~rad. Operating with mV amplitudes reduces this to $10^{2}$--$10^{4}$~rad, within the dynamic range of phase-resolved detection. The scalar coupling coefficient $\kappa/\hbar$ is unknown, but the functional form is what matters.

\subsection{Proposed measurement protocol}
\label{sec:protocol}

We propose sweeping the dimensionless parameter $\omega T$ over a range covering several zeroes and maxima of both $\sin(\omega T)$ and $1-\cos(\omega T)$. The experiment employs a symmetric electron biprism interferometer~\cite{hasselbach2010, matteucci1985}, or a single-electron Fabry-P\'{e}rot interferometer in the quantum Hall regime~\cite{bartolomei2025}. A coherent beam is split; along a segment $L$ on each arm, a cylindrical electrode driven by an RF source applies $\Phi_{1,2}(t)$.

The interference pattern is recorded as $\omega$ is stepped so that $\omega T$ varies from $0.2\pi$ to $10\pi$. At each frequency, the fringe position is extracted from a two-dimensional detector image. The transit time $T$ can be varied independently by adjusting beam energy.

Stability of the RF modulation and synchronization with electron arrival are critical at GHz frequencies. State-of-the-art RF sources achieve phase noise below $-100$~dBc/Hz at 1~GHz, corresponding to timing jitter below $10^{-12}$~s over nanosecond transit times, which is negligible. The primary noise source is shot noise. For a coherent field-emission source delivering $10^{3}$--$10^{4}$ electrons per second, a single fringe position measurement with $10^{-2}$~rad uncertainty requires $\sim 1$--$10$~s per frequency point. A full sweep of 50 points requires 10--20 minutes, compatible with typical interferometer stability~\cite{hasselbach2010}. Maximum discrimination occurs at $\omega T \sim 1$--$3$, where the standard phase has passed its first maximum while $1-\cos$ is rising.

The measured phase as a function of $\omega T$ is fit to
\begin{equation}
\Delta\phi(\omega T) = A \sin(\omega T) + B[1 - \cos(\omega T)],
\label{eq:fit}
\end{equation}
with $A$ and $B$ as free parameters. Here $A$ is not a free parameter: it is fixed by the known values of $e/\hbar$, the applied amplitude $\Delta\Phi_0$, and the modulation frequency $\omega$, giving $A = -e\Delta\Phi_0/(\hbar\omega)$. The coefficient $B$ (equivalently $\kappa/\hbar$) is the single free parameter characterizing the scalar coupling strength. A fit that allows both $A$ and $B$ to vary provides a consistency check: the extracted $A$ should agree with its independently known value within experimental uncertainty. For pure standard coupling ($B=0$), the first positive zero occurs at $\omega T = \pi$; as $B/A$ increases, this zero shifts to smaller values. For pure scalar coupling ($A=0$), zeros are at $\omega T = 2n\pi$, with twice the spacing of the standard case. A fit over several periods provides both coefficients and their uncertainties.

Key experimental requirements are RF voltage stability, electron beam monochromaticity to avoid smearing of $T$, and minimization of fringe fields at electrode apertures. Electron biprism interferometry with energy spreads below $0.1$~eV has been demonstrated~\cite{hasselbach2010}, giving $\Delta T / T \approx 10^{-3}$ at $100$~eV. For $\Delta T/T \approx 10^{-3}$ and $\omega T \sim 10$, the phase uncertainty is $\approx 0.01$~rad, below the threshold for resolving zero-crossing shifts. Electrodes with axial symmetry and end caps suppress fringe fields, keeping phase integral corrections below the level that would compromise functional discrimination.

\section{Discussion}
\label{sec:discussion}

\subsection{Implications of a positive or null result}
\label{sec:disc_implications}

If future measurements confirm the pure standard result---$\sin(\omega T)$ with a non-zero static phase---then the specific scalar coupling in Eq.~\eqref{eq:B_phase_final} is ruled out at the sensitivity level of the experiment. This would establish the first experimental upper bound on a Stueckelberg scalar coupling to electrons in the field-free regime. Such a bound would constrain the Wilson coefficient $g/\Lambda$, complementing collider and precision atomic searches for new scalar interactions.

If, alternatively, the measured phase contains a $1-\cos(\omega T)$ component, it would constitute evidence for a physical role of the Stueckelberg scalar $B$ beyond the Lorenz condition. This would be a genuine discovery: the first indication that the scalar mode of the electromagnetic potential, traditionally suppressed by the Lorenz condition, has observable consequences.

\subsection{Compatibility with precision tests}
\label{sec:disc_qed}

The scalar coupling vanishes in the static limit ($\Delta\phi_B = 0$ for DC potentials), so there is no correction to static precision tests such as the electron's anomalous magnetic moment~\cite{hanneke2008} or Lamb shift measurements. The coupling becomes operative only when the potential varies on the time scale of the electron's transit---a regime that existing precision tests do not probe. The gauge-invariant formulation guarantees that any new physics at scale $\Lambda$ is suppressed by powers of $\Lambda$, ensuring consistency with all low-energy observables for sufficiently large $\Lambda$.

\subsection{Relation to the Josephson and magnetic AB effects}
\label{sec:disc_josephson}

The AC Josephson effect, where a DC voltage $V$ produces a linearly growing phase difference $d\delta/dt = 2eV/\hbar$, is not in tension with the scalar coupling. In a Josephson junction, the DC voltage is sustained by steady charge accumulation at the electrodes, which acts as a source for the electric field via $\nabla\cdot\mathbf{E} = \rho/\epsilon_0$~\cite{tinkham2004}. The full relation $B = \nabla\cdot\mathbf{A} + \dot{\Phi}/c^2$ involves a non-zero $\nabla\cdot\mathbf{A}$ in the presence of currents, and the reduction $B = \dot{\Phi}/c^2$ used in the field-free AB configuration does not apply. Moreover, the Josephson relation $f = 2eV/h$ has been verified with extraordinary precision and forms the basis of the Josephson voltage standard~\cite{tinkham2004}. Any correction from the scalar coupling would manifest as a deviation from linearity in the voltage-frequency relation at the level of $\Delta V/V \sim \kappa/e \sim g/(e\Lambda c^2)$. For $\Lambda \gtrsim 1$~TeV and $g \sim \mathcal{O}(1)$, this correction is below $10^{-23}$, far below current experimental sensitivity. A complete treatment of the Josephson effect including the scalar coupling is beyond the scope of this work.

The magnetic AB effect is unaffected to leading order. Under $B = \partial_\mu \tilde{A}^\mu$, a static magnetic potential with $\Phi = 0$ and $\nabla\cdot\mathbf{A} = 0$ (Coulomb gauge for an ideal solenoid) gives $B = 0$, so the scalar coupling contributes nothing. The topological protection of the magnetic AB phase persists, as confirmed by precision measurements with toroidal magnets~\cite{tonomura1986}.

\subsection{Distinguishing quantum phase from classical deflection}
\label{sec:classical}

A central experimental concern is the possibility that residual electric fields from imperfect shielding could mimic or contaminate the phase signal. A stray electric field $\delta\mathbf{E}$ exerts a Lorentz force on the electron, causing a spatial deflection $\Delta x$ of the beam centroid at the detector. For a field of magnitude $\delta E$ acting over a length $\ell_{\text{stray}}$, the deflection is $\Delta x \approx (e\delta E\,\ell_{\text{stray}}/2E)L_{\text{drift}}$, where $L_{\text{drift}}$ is the distance from the interaction region to the detector and $E$ is the electron energy. At 100~eV, a residual field of $10^{-3}$ times the applied field (corresponding to $\delta E \sim 0.1$~V/m for $\Delta\Phi_0 = 1$~V across a 1~cm tube) over $\ell_{\text{stray}} \sim 0.1$~mm produces a centroid displacement of order $5$~nm for $L_{\text{drift}} \sim 0.1$~m, far below typical detector resolution. Nevertheless, this displacement can be monitored independently: a pure quantum phase shift---whether from minimal coupling or from the scalar coupling---produces a fringe shift without centroid displacement, while a classical force produces both. By simultaneously measuring fringe position and beam centroid, the two contributions can be separated. This technique has been demonstrated in the context of the MP effect by Hilbert \emph{et al.}~\cite{hilbert2011}.

\subsection{Relation to alternative formulations and large gauge transformations}
\label{sec:alternatives}

The idea that the Lorenz scalar $B = \partial_\mu A^\mu$ may be physical has a long history. Stueckelberg~\cite{stueckelberg1938} introduced it to preserve gauge invariance in massive electrodynamics. Fock and Podolsky~\cite{fock1932, dirac1932} explored formulations with a physical longitudinal mode. Ohmura~\cite{ohmura1956} proposed a scalar electrodynamics with $\Box B = 0$. More recently, various authors have investigated theories with propagating scalar electrodynamics~\cite{modanese2017, woodside1999, modanese2023b}, including a Lagrangian formulation for gauge waves in material media~\cite{modanese2023b}. The gauge-invariant formulation used here aligns naturally with these approaches: the Stueckelberg field $\phi$ provides a systematic way to control the departure from the Lorenz condition while preserving exact gauge symmetry.

The modern understanding of gauge theories has been enriched by the study of large gauge transformations and asymptotic symmetries. When gauge transformations do not vanish at infinity, they give rise to the BMS group and to soft-photon theorems~\cite{strominger2017}. In finite volumes with time-dependent boundary conditions, the separation between physical and unphysical modes becomes more subtle. The present experiment can be interpreted as a bulk probe of whether the scalar sector, normally confined to boundary degrees of freedom, acquires measurable effects when the boundary conditions are made time-dependent. A decisive advantage of the AB configuration is its immunity to classical backgrounds. In the shielded geometry, $\nabla B = \nabla(\dot{\Phi}/c^2) = 0$, so no classical force arises from the scalar coupling. The observed fringe shift isolates the quantum phase from any classical deflection.

\section{Conclusion}
\label{sec:conclusion}

We have shown that a gauge-invariant extension of QED containing a Stueckelberg scalar field predicts a phase shift with a distinctive $1-\cos(\omega T)$ signature in the shielded electrostatic AB configuration. The scalar $B = \partial_\mu \tilde{A}^\mu$, gauge-invariant by construction, couples to electrons via a Pauli-like term $B\bar{\psi}\psi/\Lambda$. In the shielded geometry, this coupling yields a phase shift proportional to the net change in the potential difference, $\Delta\phi_B \propto \Delta\Phi(T) - \Delta\Phi(0)$, rather than to its time integral.

The two functional forms---$\sin(\omega T)$ from minimal coupling and $1-\cos(\omega T)$ from the scalar coupling---are linearly independent, and the experimental protocol we have outlined sweeps $\omega T$ and fits to $A\sin(\omega T) + B[1-\cos(\omega T)]$ to separate them cleanly. The measurement is within reach of current single-electron interferometry~\cite{bartolomei2025}.

The original electric AB configuration, with time-dependent potentials applied to shielded cylinders, has never been experimentally tested. The measurement proposed here would fill this gap, provide the first frequency-resolved map of the electric AB phase, and either detect or place the first bound on a Stueckelberg scalar coupling to electrons. In doing so, it would address a question that has lingered since 1959: is the Lorenz gauge a matter of convenience, or a matter of principle?

\end{document}